\documentclass[9pt]{IEEEtran}
\IEEEoverridecommandlockouts
\usepackage{cite}
\usepackage{amsmath,amssymb,amsfonts}
\usepackage{graphicx}
\usepackage{textcomp}
\usepackage{algpseudocode}
\usepackage{makecell}
\usepackage{adjustbox}
\usepackage{amsmath}
\usepackage{braket}
\usepackage{textcomp}
\usepackage{cleveref}
\usepackage{soul}
\usepackage{fancyvrb}
\usepackage{sidecap}
\usepackage{balance} 
\usepackage{booktabs}
\usepackage{multirow}
\usepackage{float}
\usepackage{svg}
\usepackage{setspace}
\usepackage{pgfplots}
\usepackage{subcaption} 
\usepackage{tikz}
\usepackage{multirow}
\usepackage{multicol}
\usepackage{threeparttable} 
\usepackage[linesnumbered,ruled,vlined]{algorithm2e}
\usepackage{xcolor}
\usepackage{cite}
\usepackage{amsmath,amssymb,amsfonts}
\usepackage{float}
\usepackage{svg}
\usepackage{stfloats}
\usepackage{url}
\usepackage{tikz}
\usepackage{enumitem}

\def\BibTeX{{\rm B\kern-.05em{\sc i\kern-.025em b}\kern-.08em
    T\kern-.1667em\lower.7ex\hbox{E}\kern-.125emX}}

\usepackage[compact]{titlesec}

\titlespacing\section{0pt}{0.3\baselineskip}{0.2\baselineskip}
\titlespacing\subsection{0pt}{0.2\baselineskip}{0.1\baselineskip}
\titlespacing\subsubsection{0pt}{0.15\baselineskip}{0.1\baselineskip}

\usepackage{fancyhdr}
\fancypagestyle{firstpage}{
   \fancyhf{} 
   \fancyhead[C]{To appear at $62^{\text{nd}}$ IEEE/ACM Design Automation Conference (DAC) Proceedings} 
}

\begin{document}

\title{Computational Advantage in Hybrid Quantum Neural Networks: Myth or Reality?}


\author{\IEEEauthorblockN{Muhammad Kashif \IEEEauthorrefmark{1}\IEEEauthorrefmark{2},
Alberto Marchisio\IEEEauthorrefmark{1}\IEEEauthorrefmark{2}},
Muhammad Shafique\IEEEauthorrefmark{1}\IEEEauthorrefmark{2}

\IEEEauthorblockA{\IEEEauthorrefmark{1}  eBrain Lab, Division of Engineering, New York University Abu Dhabi, PO Box 129188, Abu Dhabi, UAE}\\
\IEEEauthorblockA{\IEEEauthorrefmark{2} \normalsize Center for Quantum and Topological Systems, NYUAD Research
Institute, New York University Abu Dhabi, UAE}

Emails: \{muhammadkashif, alberto.marchisio, muhammad.shafique\}@nyu.edu
    }


\maketitle
\thispagestyle{firstpage}

\begin{abstract}
Hybrid Quantum Neural Networks (HQNNs), under the umbrella of Quantum Machine Learning (QML), have garnered significant attention due to their potential to enhance computational performance by integrating quantum layers within traditional neural network (NN) architectures. 
Despite numerous state-of-the-art applications, a fundamental question remains: \textbf{\textit{Does the inclusion of quantum layers offer any computational advantage over purely classical models? If yes/no, how and why?}}

In this paper, we analyze how classical and hybrid models adapt their architectural complexity in response to increasing problem complexity. To this end, we select a multiclass classification problem and perform comprehensive benchmarking of classical models for increasing problem complexity, identifying those that optimize both accuracy and computational efficiency to establish a robust baseline for comparison.
These baseline models are then systematically compared with HQNNs by evaluating the rate of increase in floating-point operations (FLOPs) and number of parameters, providing insights into how architectural complexity scales with problem complexity in both classical and hybrid networks.
We utilize classical machines to simulate the quantum layers in HQNNs, a common practice in Noisy Intermediate-Scale Quantum (NISQ) era. 
Our analysis reveals that, as problem complexity increases, the architectural complexity of HQNNs, and consequently their FLOPs consumption, despite the simulation overhead associated with quantum layer's simulation on classical hardware, scales more efficiently ($53.1\%$ increase in FLOPs from $10$ features (low problem complexity) to $110$ features (high problem complexity)), compared to classical networks ($88.1\%$). 
%
%
%
%
Moreover, as the problem complexity increases, classical networks consistently exhibit a need for a larger number of parameters to accommodate the increasing problem complexity. Additionally, the rate of increase in number of parameters is also slower in HQNNs ($81.4\%$) than classical NNs ($88.5\%$).
These findings suggest that HQNNs provide a more scalable and resource-efficient solution, positioning them as a promising alternative for tackling complex computational problems.

\end{abstract}


\begin{spacing}{0.965}
\section{Introduction}
\begin{figure}
    \centering
    \includegraphics[scale=0.31]{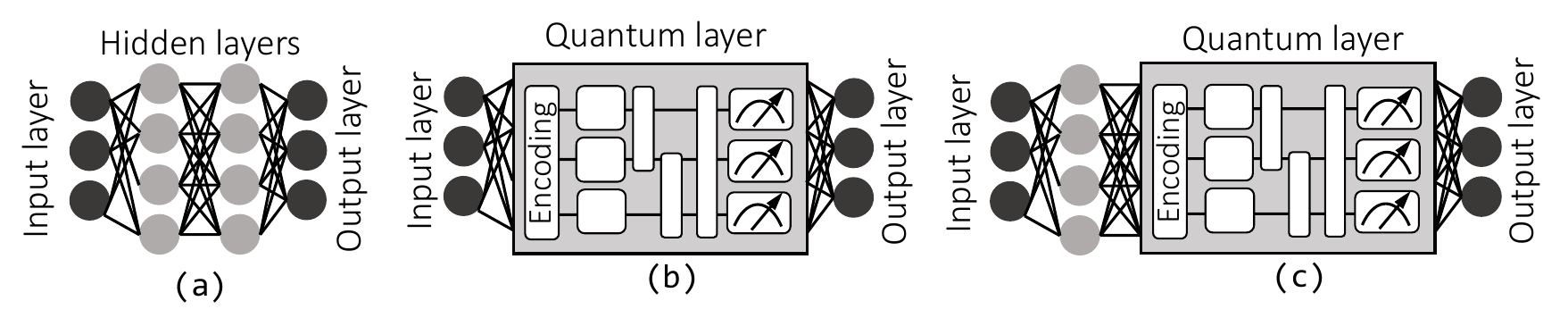}
    \vspace{-15pt}
    \caption{\footnotesize Illustration of (a) classical NN, (b) HQNN with only quantum hidden layer and (c) HQNN with classical and quantum hidden layers. }
    \label{fig:HQNN_exp}
\end{figure}

Quantum Machine Learning (QML) is an emerging research area that integrates quantum computing with machine learning (ML) techniques, aiming to enhance computational power and efficiency in processing complex datasets \cite{Schuld:2015,zhang:2020,peral:2024,Zaman_2024arxiv_QMLSurvey}. 
Quantum Neural Networks (QNNs) are a core component of QML, employing quantum circuits/layers to mimic the structure and function of classical neural networks (NNs)\cite{abbas:2021, Kashif:2023_impact,kashif:2023_param_init_classical}.
Hybrid Quantum Neural Networks (HQNNs) advance this concept by combining quantum and classical NN, leveraging the strengths of both quantum and classical computation\cite{kashif:2021,zaman:2024_impact, Kashif_2024_Investigating_noise}. 
HQNNs are particularly advantageous in the Noisy Intermediate-Scale Quantum (NISQ) era, where they can capitalize on quantum capabilities for specific sub-tasks while relying on classical resources to manage error-prone operations \cite{kashif_unified,bokhan:2022}. 
In a typical HQNN architecture, one or more hidden layers of a classical NN are replaced by a trainable quantum layer \cite{kashif:2022a,kashif:2024HQNET}, as shown in Fig. \ref{fig:HQNN_exp}. 

A growing body of state-of-the-art research employs HQNNs across a wide range of applications \cite{domingo:2023,dong:2023,domingo:2023a,kashif2024resqnets,innan:2024,kaseb:2024,chen:2024,wang:2024,paquet:2024}. Despite the notable computational speedups demonstrated by quantum computing in various domains \cite{Arute:2019,zhong:2020,Wu:2021,Madsen:2022}, no prior study has systematically investigated whether the integration of \emph{quantum} components in NNs offers any advantages in learning performance or computational efficiency\cite{bowles:2024}. Additionally, demonstrating quantum advantage for QML is challenging due to several reasons, as critically discussed in~\cite{Schuld:2022}.

\subsection{Open Research Questions}
Comparing classical and quantum models is not straight-forward due to their fundamentally different computational paradigms. 
In the NISQ era, where quantum computers are limited by noise and scale, most research relies on simulating quantum algorithms using classical hardware\footnote{List of quantum simulators (Accessed 15-11-2024): \url{https://www.quantiki.org/wiki/list-qc-simulators}}\cite{bowles:2024}, which is computationally intensive due to the exponential scaling of quantum states \cite{Preskill_2018}.
\textit{By examining the increase in architectural complexity (required to address growing problem complexity) and the corresponding increase in computational demands of QML models simulated on classical machines, potential quantum advantages over classical models can be evaluated.} 
The argument is that, if QML models exhibit superior performance despite the overhead associated with simulating quantum circuits on classical hardware, this suggests that the observed superiority/advantage is likely inherent to the quantum nature of the algorithms and could become more significant with the deployment of universal fault-tolerant quantum computers.
%
\textit{This approach provides valuable insights into the scalability and impact of quantum computing, offering a pragmatic pathway to evaluate its future role in solving complex problems beyond classical capabilities.}

To this end, recently, there have been some efforts on benchmarking QML algorithms, primarily focusing on analyzing which algorithms perform better in specific scenarios \cite{bowles:2024,schnabel:2024}. However, the question of whether incorporating the quantum component in NNs framework provides any advantage remains unexplored and is an \emph{open and important} research problem. \textbf{In this paper, we raise several key questions to investigate this gap:}

\begin{itemize}[leftmargin=*]
    \item \textbf{Q1:} \textit{What metrics will be appropriate to evaluate and compare the computational complexity of classical and hybrid networks?}

    \item \textbf{Q2:} \textit{Does the quantum part in HQNNs add anything qualitatively different or important?}
    
    \item \textbf{Q3:} \textit{If the quantum part gives any advantage in HQNNs, and can we demonstrate if it actually is computational advantage?}
\end{itemize}

\subsection{Our Contributions}
We attempt to answer the above questions through a rigorous empirical analysis. \textbf{Our main contributions are:}

\begin{figure}[htbp]
    \centering
    \includegraphics[scale=0.4]{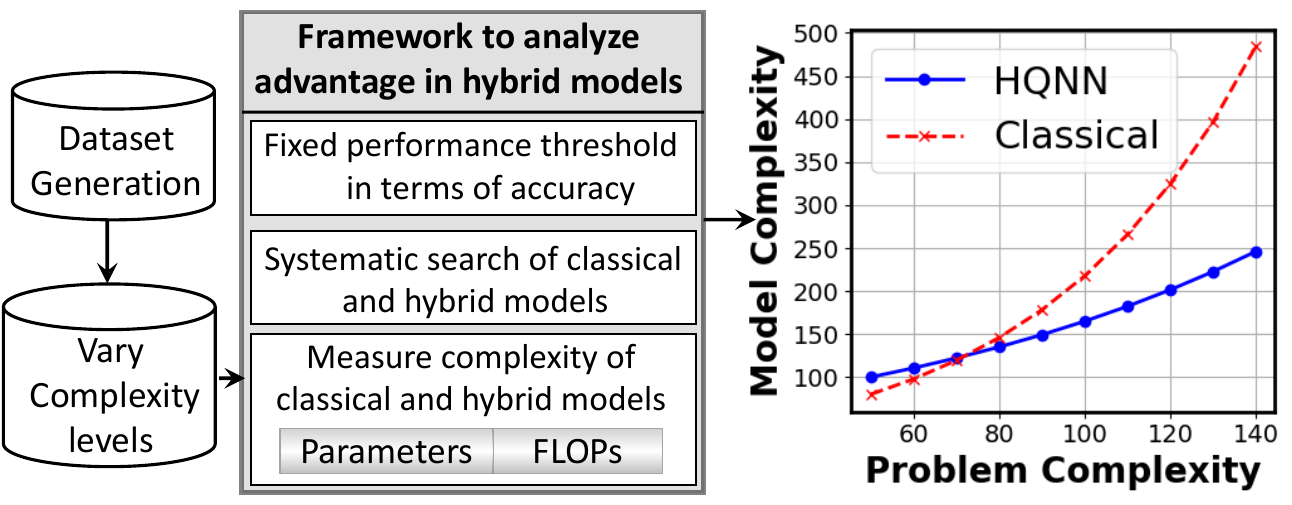}
    \caption{\footnotesize An overview of our contributions. }
    \label{fig:cont_syn_data}
\end{figure}

\begin{itemize}[leftmargin=*]
    \item We propose a systematic methodology for comparing the computational complexity of hybrid and classical networks. For this purpose we generate a synthetic dataset in such a way that we can increase the complexity of the problem by increasing the number of features. {Our novel investigation unleash how both classical and hybrid networks adapt to the complexity of the problem.} (\textbf{Section~\ref{sec:methodology}}) 

    \item We perform grid search to identify the classical models that meet a predefined accuracy threshold across varying levels of problem complexity to serve as benchmarks. We then identify the most efficient hybrid network architectures (models that achieve the predefined accuracy threshold) through a similar grid search process. (\textbf{Section \ref{sec:classical_model_search} and \ref{sec:hybrid_model_search}})

    \item A key finding from our analysis is that \textit{as the problem complexity increases, classical NNs require more sophisticated and resource-intensive architectures to reach a certain accuracy, resulting in higher FLOPs consumption}. In contrast, \textit{HQNNs demonstrate superior scalability and efficiency}. A well-designed hybrid architecture, particularly with a more \textit{expressive quantum layer}, remains largely unaffected by the increasing complexity of the problem. An architecture that effectively solves a simpler problem can often be adapted, to solve a more complex problem. (\textbf{Section~\ref{sec:results}})

    \item Another important finding of our analysis is that \textit{classical models consistently require a greater number of parameters across all levels of problem complexity, even at lower complexity levels}. In contrast, \textit{HQNNs generally need fewer parameters, highlighting their efficiency in parameter utilization.} (\textbf{Section \ref{sec:results}}) 
    
\end{itemize}

\begin{figure*}[b]
    \centering
    \includegraphics[scale=0.335]{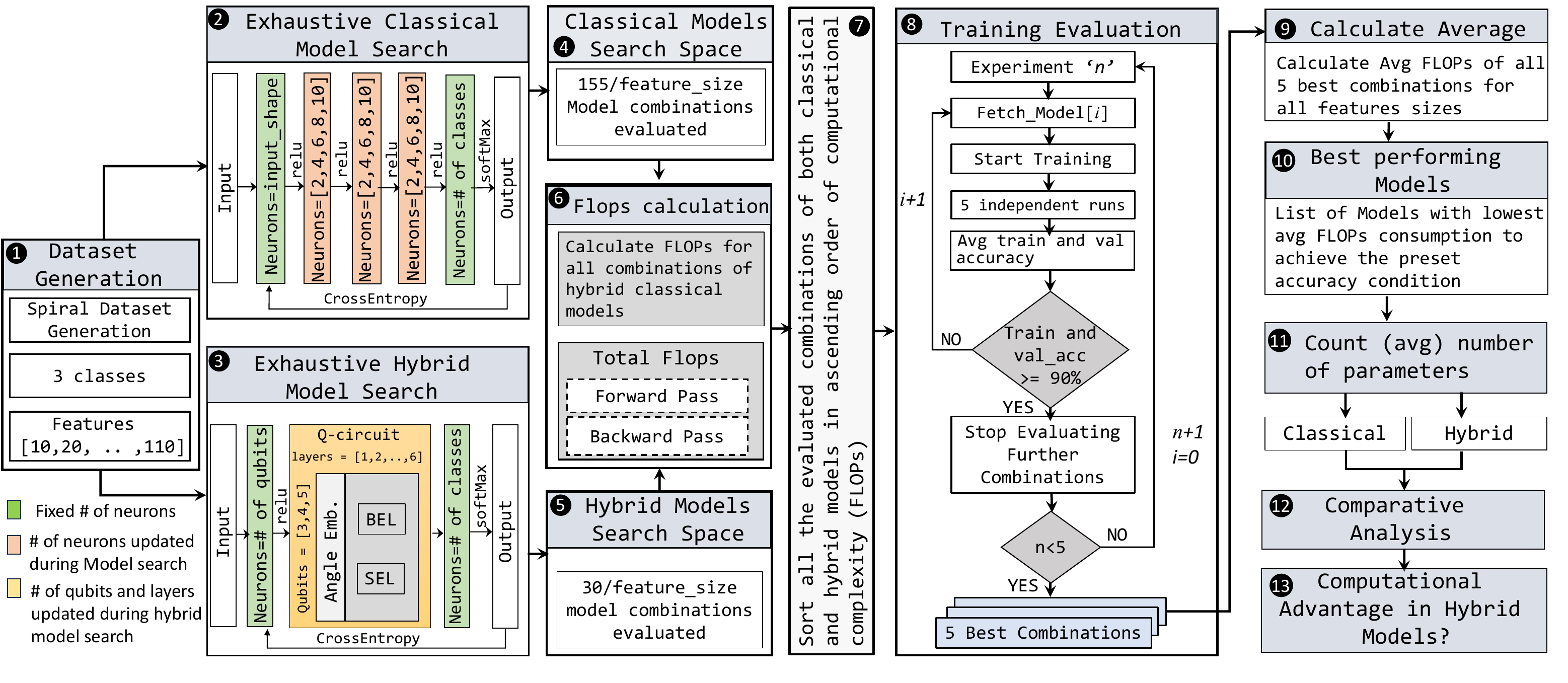}
    \vspace{-10pt}
    \caption{\footnotesize Detailed Methodology. 
    We analyze how classical and hybrid models scale with problem complexity. We generate synthetic dataset which allows to progressively increase the problem complexity. We then search for classical and hybrid models using grid search approach and compare the FLOPs and parameters. To account for randomness in NNs training, the results are averaged over 5 experiments. Small black boxes show the order of methodology steps.}
    \label{fig:methodology}
\end{figure*}

{\textbf{\textit{Summary of key results:}}}
This study demonstrates the superior scalability of HQNNs compared to classical NNs as problem complexity increases. In our analysis, we find that despite simulation overhead, HQNNs with expressive quantum layers require approximately $7.5\%$ fewer FLOPs and $42.3\%$ fewer parameters to solve complex problems ($110$ features in our case) than classical networks. 
Additionally, HQNNs scale more efficiently, with a $53.1\%$ increase in FLOPs consumption while going from simple to complex tasks, compared to $88.5\%$ increase in FLOPs for classical models. Similarly, the rate of increase in number of parameters for classical models is $88.5\%$, while HQNNs show a slower rate of increase in parameters that accounts to $81.4\%$.



\section{Complexity of Neural Networks}
There is no single, definitive measure for assessing the NN's complexity, however, several approximate metrics exist\cite{shalev:2014,Lu:2017,Lee:2020,lech:2022, holstermann:2023,ehrlich:2023}. Therefore, we argue that relying on a single measure may not provide a comprehensive understanding of the network's complexity. Hence, it is advisable to use multiple methods to better estimate the network's complexity for a given task. We use FLOPs and the number of parameters as indicators to assess the network's complexity.

\subsection{Floating Point Operations (FLOPs)}
FLOPs are a widely recognized measure of computational complexity in NNs \cite{Tan:2019,Guo:2021,hsia:2021,Chen:2023}. FLOPs provide an estimate of the total number of arithmetic operations required to execute a model on a given input. This metric serves as a proxy for understanding the computational resources, such as processing power and energy consumption that are needed to train and evaluate NNs. By quantifying FLOPs, one can objectively compare the efficiency of different architectures, enabling more informed decisions about model design and scalability.

\subsection{Number of Parameters} 
The number of parameters represent a total number of trainable weights in a NN and is often used as a measure of model complexity, with larger networks (with more parameters) generally being more expressive and capable of capturing complex data patterns \cite{Goodfellow:2016}. In NNs, the parameter count typically scales with the number of layers and neurons per layer, and is hence useful for evaluating model complexity \cite{bianchini:2014}.


\section{Our Methodology} \label{sec:methodology}
In this paper, we argue that two models are comparable if they achieve similar performance (accuracy) on a given task, regardless of differences in their size. Accordingly, we set an accuracy threshold of $\geq 90\%$ for both training and validation, and perform an extensive grid search for models that meet this condition regardless of their size.
The detailed methodology is shown in Fig. \ref{fig:methodology}. Below, we explain different steps of our methodology in detail.


\subsection{Synthetic Dataset Generation} \label{sec:data_gen}
\begin{figure}[htbp]
    \centering
    \includegraphics[scale=0.35]{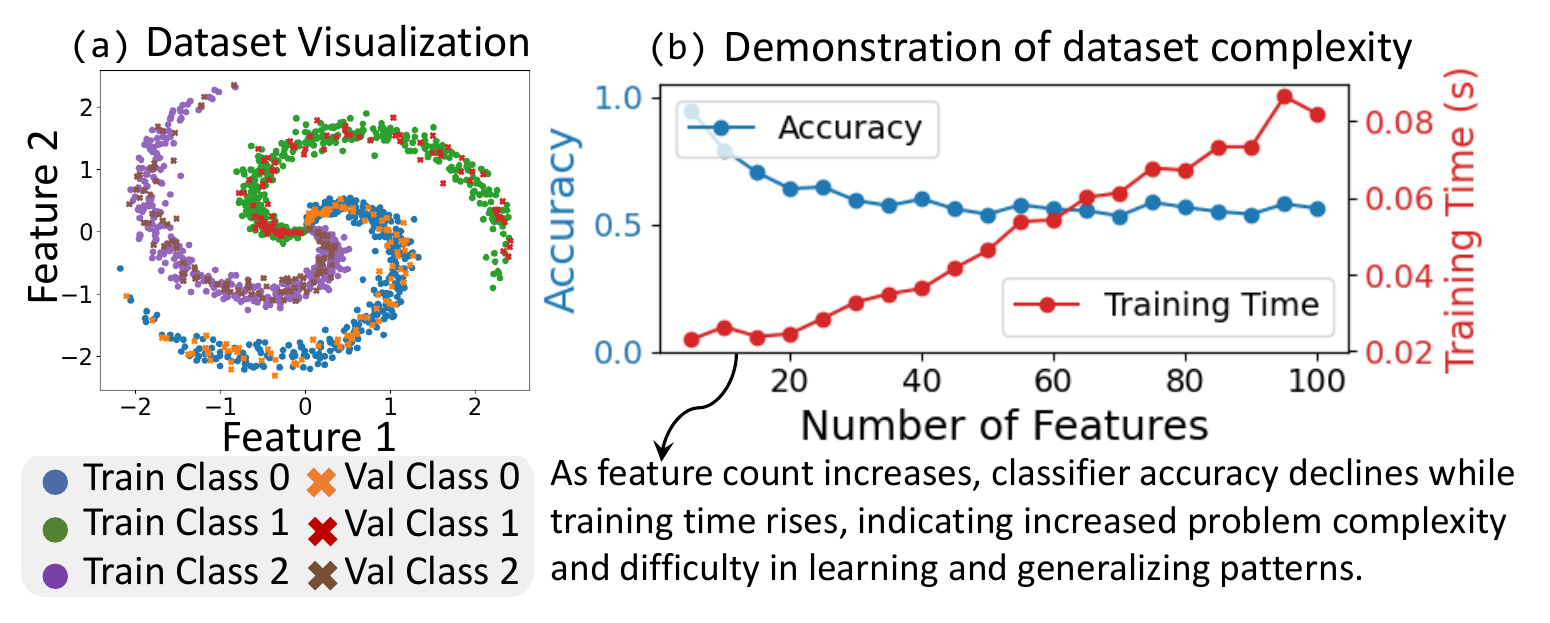}
    \caption{\footnotesize (a) Visualization of first two features of generated dataset, and (b) Demonstration of increasing problem complexity.}
    \label{fig:dataset_complexity}
\end{figure}

We generate synthetic datasets as these datasets play a crucial role in benchmarking ML models, offering precise control to systematically increase task complexity and evaluate model performance under varying conditions\cite{liu:2021, patryk:2022, bowles:2024}.
We generate a spiral dataset with $1500$ data points distributed across $3$ classes. The dataset's core structure is a spiral, as shown in Fig. \ref{fig:dataset_complexity}(a) with each class represented by a distinct arm, offering a challenging yet learnable pattern for models.
To investigate the effect of task complexity, we maintain a fixed number of data points and classes while systematically increasing complexity by augmenting the number of features. The increase in feature dimensions is accompanied by a corresponding increase in noise, further enhancing the complexity of the dataset. Specifically, the additional features introduce subtle variations through non-linear transformations of the existing features. Controlled noise is introduced, scaled with the number of features $(noise=0.1+0.003 \times num\_features)$, ensuring a progressive rise in task difficulty.
We utilize feature sizes ranging from $10$ to $110$ in increments of $10$. From this point onward, we refer to feature size as complexity level, with the associated noise implicitly applied unless explicitly stated otherwise. This incremental increase in feature dimensions results in a progressively challenging classification task, as illustrated in Fig. \ref{fig:dataset_complexity}(b).

\subsection{Exhaustive Classical Model Search} \label{sec:classical_model_search}
We use grid search algorithm to systematically search for the models that reach the preset accuracy condition ($\geq90\%$).


\textbf{\textit{Working of Grid Search Algorithm}:}
For $n$ layers, and $m$ neuron options, our grid search algorithm evaluates all the possible combinations based on formula: $ \text{Total no. of combinations} = m \frac{m^n-1}{m-1}$.
For instance, if $m = [2,3]$ (i.e., there can be 2 or 3 neurons in each layer, resulting in neuron option $m=2$), and $n = 2$, this means we can have minimum $1$ and maximum $2$ layers. The neurons in each layer can be either $1$ or $2$. Using the general formula, there are $6$ possible combinations, i.e., $[2], [3], [2,2], [2,3], [3,2], [3,3]$.

\textbf{\textit{Classical Model Search Space:}}
We restrict the classical models to have a maximum of $n = 3$ layers, with the number of neurons in each layer chosen from the set $m=\{2,4,6,8,10\}$ ($m=5$), resulting in a search space of total $155$ model combinations for classical models for each complexity level.

\subsection{Exhaustive Hybrid Model Search}\label{sec:hybrid_model_search}
For hybrid models, the number of neurons in the first and last classical layers are fixed. The number of neurons in the input layer are equal to the number of qubits because we use angle encoding that requires one qubit to encode one feature\cite{Larose:2020}. The neurons in the output layer are equal to the number of classes ($3$) in the dataset. 

\textit{\textbf{Hybrid Model Search Space:}}
During the hybrid model search only the quantum layers are varied. We use $[3, 4, 5]$ qubits quantum layers, and for each qubit size, quantum layers of depth $[1, 2, 3, ...,10]$ are tested, yielding $30$ model combinations per feature size. 

\begin{figure}
    \centering
    \includegraphics[scale=0.35]{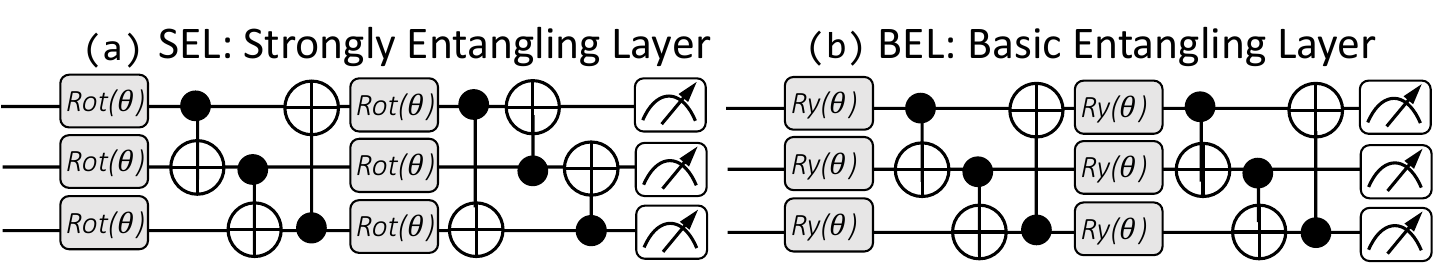}
    \caption{\footnotesize Example of (a) SEL and (b) BEL quantum layer design, both having $3$ qubits and depth of $2$ layers. Same structure is repeated for more layers.}
    \label{fig:BEL_SEL}
\end{figure}

\begin{figure*}[t]
    \centering
    \includegraphics[scale=0.48]{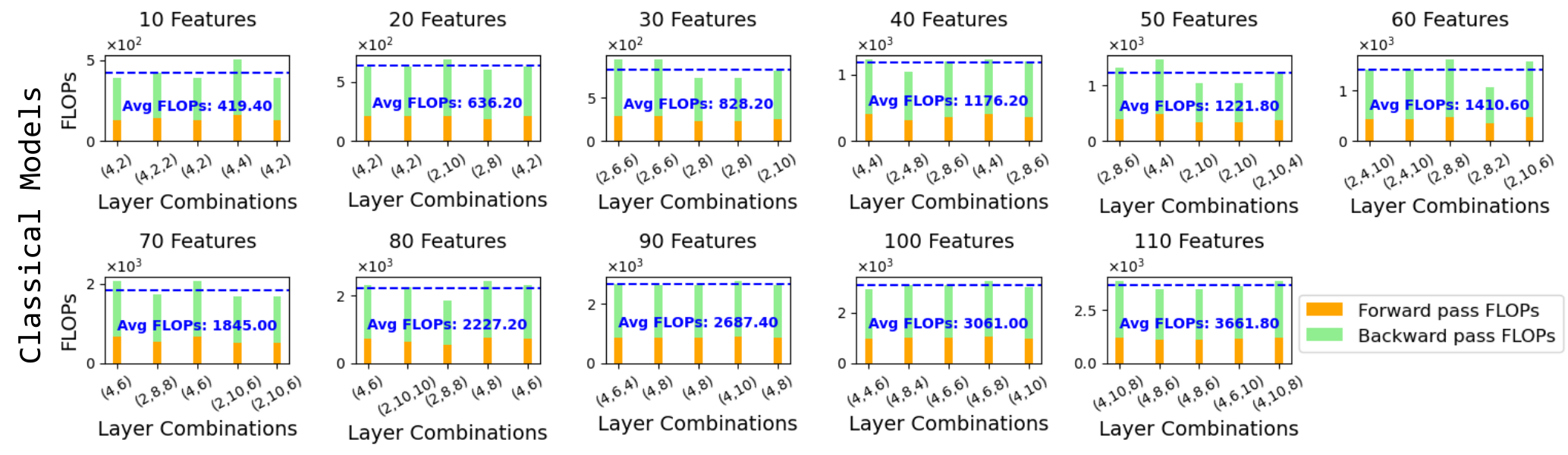}
    \caption{\footnotesize FLOPs consumption of best-performing classical models for different complexity levels of the problem. Each subplot displays the FLOPs consumption of the top five performing models from five independent runs of model search, corresponding to different feature sizes.}
    \label{fig:avg_flops_classical}
\end{figure*}

\textbf{\textit{Quantum Layer Design:}}
For the underlying quantum layers in HQNNs, we use two different types designs which are separately evaluated: the Strongly Entangling Layer (SEL) and the Basic Entangling Layer (BEL), as shown in Fig. \ref{fig:BEL_SEL}. For both layer types, each qubit number is experimented with all the layer depths, to identify the optimal configuration, i.e., that meets the preset accuracy threshold of $\geq90\%$.


\subsection{FLOPs Calculation}
We use built-in functions from Tensorflow to compute the FLOPs for both and forward and backward pass. The procedure to compute FLOPs involves converting the \texttt{TensorFlow/Keras} models (both classical and HQNNs\footnote{We convert quantum layers in HQNNs into Tensorflow/keras layer using pennylane, more details at: \url{https://pennylane.ai/qml/demos/tutorial_qnn_module_tf}}) into a concrete function using TensorFlow's \texttt{convert\_variables\_to\_constants\_v2} utility. The frozen computation graph of the model is then analyzed using the \texttt{TensorFlow Profiler} with options set to compute the total number of floating-point operations.

For the forward pass, the profiler directly computes FLOPs based on the operations in the model graph. For the backward pass, we define a dummy loss function and use TensorFlow's \texttt{GradientTape} to compute gradients, which are also converted to a frozen computation graph. FLOPs for the backward pass are then calculated in a similar manner. The total FLOPs are obtained by summing the FLOPs of the forward and backward passes, providing a comprehensive measure of the model's computational complexity.



\subsection{Models Sorting based on FLOPs consumption}
Once the FLOPs are calculated, the model combinations are sorted in ascending order based on computational complexity (higher FLOPs means complex model and vice versa).
This sorting ensures that we avoid training all models unnecessarily as the first model in the list that satisfies the accuracy condition is the model with the lowest computational complexity.


\begin{figure*}[b]
    \centering
    \includegraphics[scale=0.48]{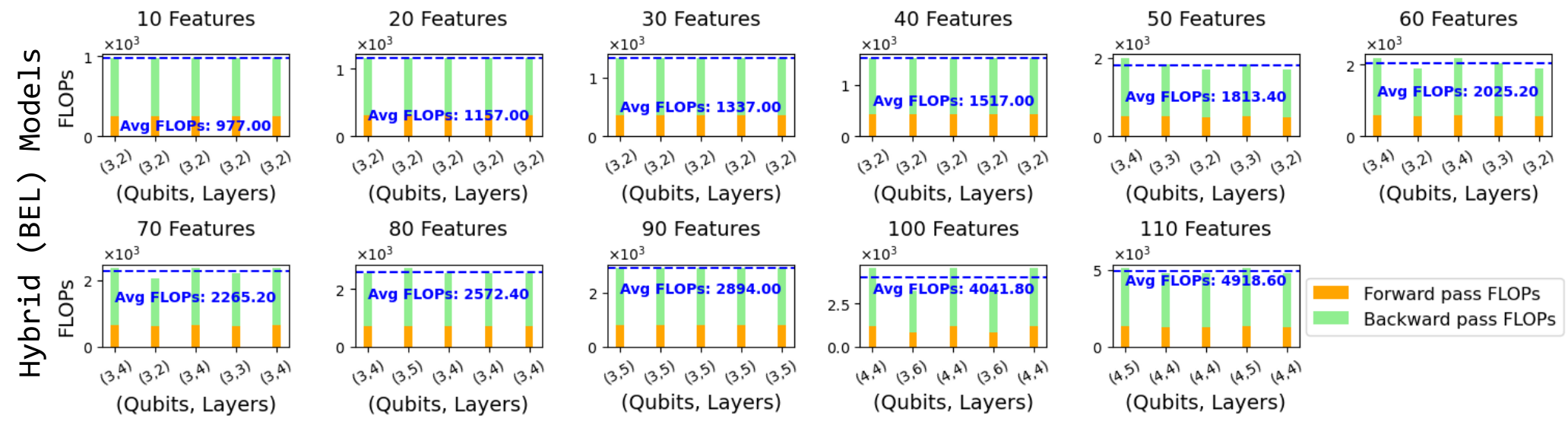}
    \caption{\footnotesize FLOPs consumption of best-performing hybrid (BEL) models for different complexity levels of the problem. Each subplot displays the FLOPs consumption of the top five performing models from five independent runs of model search, corresponding to different feature sizes.}
    \label{fig:avg_flops_BEL}
\end{figure*}

\begin{figure*}[h]
    \centering
    \includegraphics[scale=0.48]{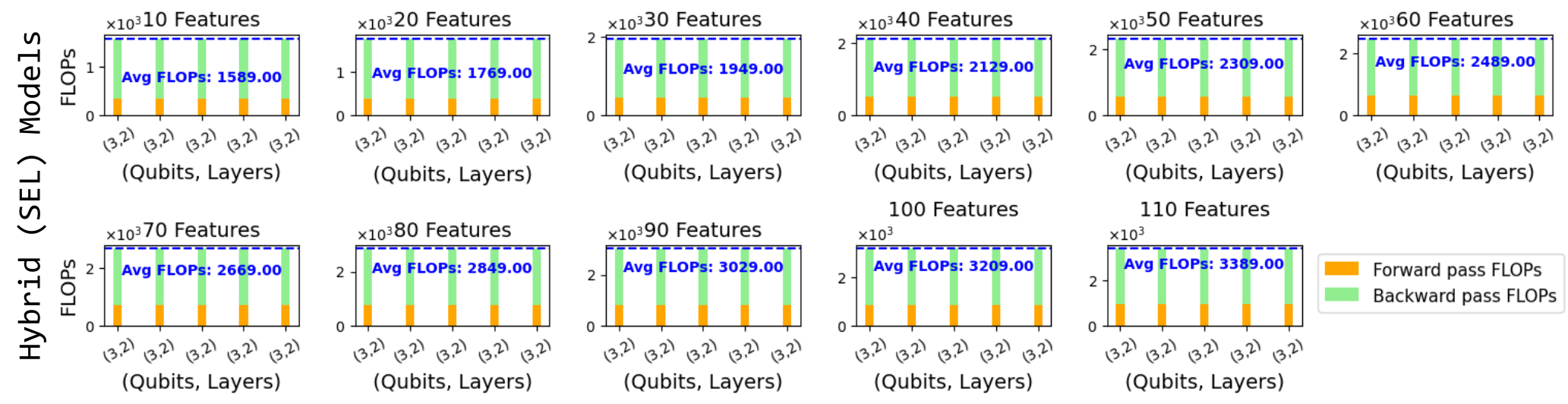}
    \caption{\footnotesize FLOPs consumption of best-performing hybrid (SEL) models for different complexity levels of the problem. Each subplot displays the FLOPs consumption of the top five performing models from five independent runs of model search, corresponding to different feature sizes.}
    \label{fig:avg_flops_SEL}
\end{figure*}
\subsection{Models Training} \label{sec:method_model_training}
The sorted list of models are trained sequentially for various levels of problem complexity. To account for the inherent randomness in parameter initialization and to ensure the statistical reliability of the results, each model configuration is evaluated over five independent runs. This approach mitigates the effects of stochastic variations and provides a more robust assessment of model performance across different initialization instances. Each run involves training for $100$ epochs, and the highest training and validation accuracy across epochs is recorded. The maximum accuracies from all runs are averaged to provide a robust performance measure for each model combination for a given complexity level.
If any model achieves an average training and validation accuracy of $\geq 90\%$, further evaluation of additional model combinations for that complexity level is stopped. 
This entire process is repeated five times to address the stochastic nature of NNs training, where different model configurations might result as optimal at different time instances. By repeating the procedure, a more accurate estimate of the average complexity required to solve the problem is obtained. Once five top-performing combinations are identified for a given complexity level, the evaluation continues with the next complexity level.



\subsection{Models with Lowest Flops Consumption}
At the end of the experimentation process, the average computational complexity required by both classical and hybrid networks is obtained for varying levels of problem complexity. These average complexities were then compared to determine whether hybrid networks provide any computational advantage over classical networks as the complexity of the problem increases. This comparative analysis is crucial in evaluating the scalability and efficiency of hybrid models, particularly in scenarios where the problem's complexity poses significant computational challenges for classical approaches.


\section{Results and Discussion} \label{sec:results}
For the experiments, we utilized PennyLane, a Python-based framework designed for differentiable quantum programming \cite{Bergholm:2018}. During the training phase, the learning rate was set to $0.001$, and a batch size of $8$ was employed for both classical and hybrid models, with a total of $100$ training epochs. 

\subsection{Classical Models: Analysis of FLOPs Consumption}
In this section, we present our findings on the computational complexity (in terms of FLOPs consumption) required by classical models to solve the problem under predefined accuracy conditions, as shown in Fig. \ref{fig:avg_flops_classical}. By varying the problem complexity (increasing features and noise, see Section \ref{sec:data_gen}), we evaluate the scalability and computational efficiency of classical models.
%
%
%
We observe that at lower complexity, simpler models achieve the desired accuracy. However, as the problem complexity increases (e.g., up to $110$ and associated high noise level), more sophisticated models, i.e., with additional layers and neurons are needed, resulting in a significant increase in FLOPs and hardware resources.


\subsection{Hybrid Models: Analysis of FLOPs Consumption}

We now present the results illustrating how both the hybrid models adapt to increasing problem complexity.

\textbf{\textit{Discussion for BEL-based HQNNs:}}
The results in Fig. \ref{fig:avg_flops_BEL} show the computational complexity of the BEL-based hybrid model for varying problem complexities. For problems with up to $40$ features, the same quantum circuit architecture ($3$ qubits, $2$ layers) is sufficient. The increase in FLOPs is mainly due to the increase in classical input layer size as feature count grows, while the quantum layer remains unchanged in depth (layer repetitions) and width (no. of qubits). However, further increase in problem complexity necessitates making the underlying quantum layer more expressive to effectively handle the increased complexity of the problem. This involves adding more qubits and increase quantum layer's depth. These changes in the architecture lead to a significant rise in FLOPs, reflecting an increase in the overall computational demands and greater hardware resource consumption. 
%
%

%
\textbf{\textit{Discussion for SEL-based HQNNs:}}
The SEL quantum layer has a more intricate entanglement design than the BEL, enhancing its expressiveness and ability to capture complex patterns, as shown in Fig. \ref{fig:BEL_SEL}(b). Unlike BEL, the SEL structure remains fixed at $3$ qubits and $2$ layers, effectively solving problems with up to $110$ features without altering the quantum architecture, as shown in Fig. \ref{fig:avg_flops_SEL}. The increase in FLOPs is attributed to the classical component's growth, particularly in the input layer to accomodate higher feature size, while the quantum layer remains unchanged, providing an efficient framework for handling complex tasks without additional quantum resources.

 \begin{figure*}[h]
    \centering
    \includegraphics[scale=0.48]{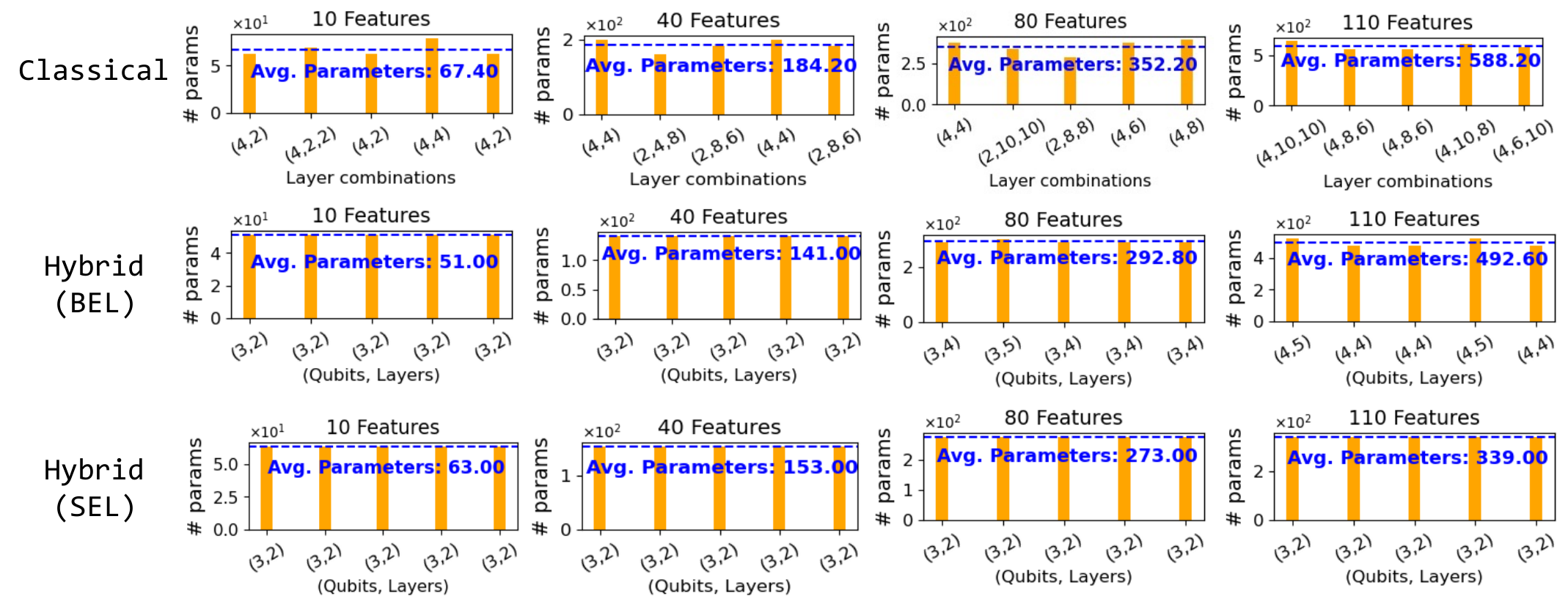}
    \caption{\footnotesize Number for parameters for five top-performing classical models (top pannel), Hybrid (BEL) Models (middle pannel) and Hybrid (SEL) Models (bottom pannel) for different level of problem complexity.}
    \label{fig:avg_params_all}
\end{figure*}
%
\textit{To make the argument tighter, we count the number of parameters of the best-performing models for both classical and hybrid approaches. This analysis is crucial, as it is possible that hybrid models have more parameters. This would potentially explain why their complexity in terms of rate of increase in FLOPs consumption does not increase as rapidly as classical models—they may be sufficiently expressive from the start. Due to page limitations, we present parameter counts for selective feature sizes ($10, 40, 80, 110$) for both classical and hybrid models. }

\subsection{Classical Models: Analysis of \# of Parameters}
For classical models, we observed a substantial increase in the number of trainable parameters as problem complexity is increased. This is primarily due to the need for more sophisticated architectures, such as additional layers and neurons, to maintain the desired accuracy with for increased complexity (Fig. \ref{fig:avg_params_all}, top panel). The trend suggests that classical models require enhanced expressiveness to manage the increased input dimensionality and noise. The increase in parameters correlates closely with the rise in FLOPs, further reinforcing the notion that classical models demand greater computational resources as problem complexity increases.

%
%

\subsection{Hybrid Models: Analysis of \# of Parameters}
For BEL-based hybrid models, the model complexity remains constant for problem complexity up to $40$ features, with an average parameter count lower than that of classical models. Beyond $40$ features, however, an increase in the number of qubits and quantum circuit depth is necessary, leading to a moderate rise in parameter count (Fig. \ref{fig:avg_params_all}, middle panel). This increase is mainly due to the larger input layer resulting from the bigger feature size. Despite this, BEL-based hybrid models maintain a lower parameter count compared to their classical counterparts.
In contrast, hybrid models with the SEL quantum layer exhibit minimal changes in parameter count across all complexity levels. The same quantum layer configuration is sufficient regardless of the problem complexity, with only a slight increase in parameters due to additional neurons in the input layer. Overall, SEL-based hybrid models demonstrate superior parameter efficiency compared to both classical and BEL-based hybrid models.
%
%

\subsection{Classical Vs. Hybrid - Comparative Analysis of FLOPs and \# of Parameters}
We now compare the classical and hybrid models in terms of FLOPs consumption and the number of parameters. The comparative analysis is aimed to provide a clear picture of how the complexity of classical and hybrid models scale with problem complexity. As discussed in Section \ref{sec:method_model_training}, five independent experiments are conducted for each complexity level to identify the best-performing models. For this comparative analysis, we select the smallest model from the set of five best-performing configurations for both classical and hybrid approaches.



\begin{figure}[h]
    \centering
    \includegraphics[width=\linewidth]{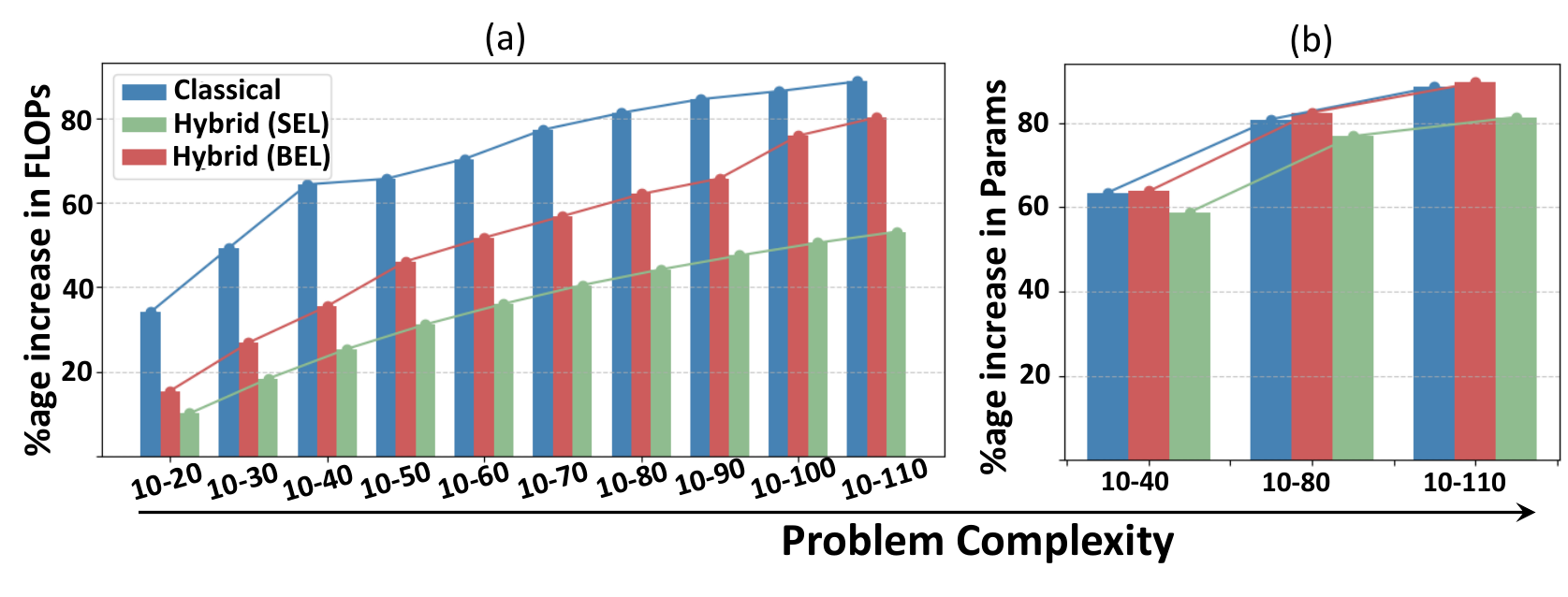}
    \vspace{-8pt}
    \caption{\footnotesize Comparison of rate of increase in FLOPs and \# of parameters of classical and HQNNs with increasing probem complexity. As the problem complexity increases, the rate of increase in FLOPs is more pronounced in classical models than hybrid models. Similarly, the rate of increase of parameter is also better in HQNNs especially with strongly entangling layers. }
    \label{fig:flops_params_comparison}
\end{figure}

\paragraph{Comparison of FLOPs Consumption}
FLOPs comparison of best-performing classical and both the hybrid models, as a function of the problem complexity is presented in Fig. \ref{fig:flops_params_comparison}(a). The results demonstrate how these models scale in terms of computational complexity as the problem complexity increases.
%
The classical model exhibited a consistent linear rate of 
increase in FLOPs consumption with increasing problem complexity. Specifically, there was an absolute increase of $3285$ FLOPs (obtained by subtracting the avg. FLOPs of $10$ features from FLOPs of $110$ features in Fig. \ref{fig:avg_flops_classical}), representing an $88.5\%$ rise in computational demands when scaling from $10$ to $110$ features.

BEL-based hybrid models also exhibit a linear rate of increase in FLOPs but at a slightly lower rate than classical models, across the same complexity levels, with an absolute rise of $3941.6$ FLOPs (obtained by subtracting the avg. FLOPs of $10$ features from FLOPs of $110$ features in Fig. \ref{fig:avg_flops_BEL}). representing an increase of $80.13\%$ computational demands when scaling from $10$ to $110$ features..
%
%
SEL-based hybrid models also showed a linear increase in computational demand, but at a significantly lower rate than both classical models and BEL-based hybrid models. The absolute increase was only $1800$ FLOPs (obtained by subtracting the avg. FLOPs of $10$ features from FLOPs of $110$ features in Fig. \ref{fig:avg_flops_SEL}), when the number of features increased from $10$ to $110$, reflecting $53.1\%$ rate of increase in FLOPs. This is substantially lower than the rates observed for classical and BEL-based models, despite the overhead of simulating quantum layers on classical hardware.

An important observation is that the rate of increase in computational demands, measured in terms of FLOPs consumption, is significantly lower in hybrid models compared to classical models. This underscores the superior adaptability of HQNNs to increasing problem complexity, particularly in HQNNs with more expressive underlying quantum layers, such as those utilizing SEL architectures.

\paragraph{Comparison of \# of Parameters}
Fig. \ref{fig:flops_params_comparison}(b) compares the average number of parameters for classical and both hybrid models as a function of problem complexity.
Classical models exhibit a steady rise in parameter count with increasing feature size, reflecting the need for greater number of parameters and expressiveness to maintain a certain performance as problem complexity grows. 
Specifically, an absolute increase of $520.8$ parameters is observed (obtained by subtracting the avg. parameters of $10$ features from avg. params of $110$ features in Fig. \ref{fig:avg_params_all}(top panel)), corresponding to an $88.5\%$ rate of increase in parameter count, as the problem complexity increases from $10$ to $110$ features.

BEL-based hybrid models also exhibit an increase in parameter count with growing problem complexity. In particular, an absolute increase of $441$ parameters (obtained by subtracting the avg. parameters of $10$ features from avg. params of $110$ features in Fig. \ref{fig:avg_params_all}(middle panel)), at a rate of $89.6\%$ is observed, which is very close to the rate of increase of parameters in classical models. 
SEL-based hybrid models demonstrate the slowest parameter growth, with incrementally increased problem complexity. While going from low to high problem complexity an absolute increase of $276$ parameters (obtained by subtracting the avg. parameters of $10$ features from avg. params of $110$ features in Fig. \ref{fig:avg_params_all}(bottom panel)) at a rate of $81.4\%$ is observed, which is significantly lower than both classical and BEL-based hybrid models. 

These results indicate that the rate of increase in parameter count as a function of problem complexity aligns closely with the rate of increase in FLOPs consumption. Notably, HQNNs employing more expressive quantum layers, such as SEL, demonstrate exceptional adaptability to increasing problem complexity, requiring minimal modifications to their architectural complexity. This observation suggests a potential computational advantage of incorporating quantum components into neural networks, highlighting the inherent benefits of quantumness in addressing complex problems.






\subsection{Ablation Study: Determining the FLOPs of Quantum Layers}
The HQNN framework incorporates classical layers for input preprocessing and postprocessing, as well as data encoding routines, which are computationally intensive when exploiting classical machines to run HQNN models. Once the universal fault-tolerant quantum computers and the development of proper quantum backpropagation techniques for QML algorithms would be developed, the reliance on classical layers and classical optimization routines is expected to diminish. Furthermore, the availability of quantum-native datasets would eliminate the need for data encoding, as the data would inherently exist in a quantum-compatible format. 
Therefore, we now compute the FLOPs consumption for classical and encoding components of HQNNs, to estimate the FLOPs consumption of actual trainable quantum layers. Table~\ref{tab:flops_segregation} provides a detailed breakdown. 
Understandably, as feature size increases, total FLOPs consumption increases for both hybrid models, due to the higher input dimensionality. The hybrid (BEL) models exhibit a more rapid increase, indicating greater sensitivity to input complexity.
%
%
%
\begin{table}[t!]
\footnotesize
\centering
\caption{\footnotesize Ablation study Results: Breakdown of FLOPs consumption for different across the different stages of the hybrid networks (in FLOPs). FS=feature size, BC=best combination, TF=total flops, Enc=Encoding, CL=classical layers, QL=quantum layer}
\begin{adjustbox}{max width=1\linewidth}
\begin{tabular}{|c|c|c|c|c|c|c|}
\hline
\textbf{Model}   & \textbf{FS/BC} & \textbf{TF} & \textbf{Enc+CL} & \textbf{CL} & \textbf{Enc} & \textbf{QL} \\ \hline
\multirow{4}{*}{\textbf{ Hybrid (BEL) }}    & 10/(3,2)   & 977   & 749    & 283    & 749-283=466  & 228  \\ \cline{2-7}
    & 40/(3,2)   & 1517  & 1289   & 823    & 1289-823=466 & 228  \\ \cline{2-7}
   & 80/(3,4)   & 2537  & 2009   & 1543   & 2009-1543=466 & 528  \\ \cline{2-7}
    & 110/(4,4)  & 4797  & 3901   & 2769   & 3901-2769=1132 & 896  \\ \hline
\multirow{4}{*}{\textbf{ Hybrid (SEL) }}     & 10/(3,2)   & 1589  & 749    & 283    & 749-283=466  & 840  \\ \cline{2-7}
    & 40/(3,2)   & 2129  & 1289   & 823    & 1289-823=466 & 840  \\ \cline{2-7}
     & 80/(3,2)   & 2849  & 2009   & 1543   & 2009-1543=466 & 840  \\ \cline{2-7}
     & 110/(3,2)  & 3389  & 2549   & 2083   & 2549-2083=466 & 840  \\ \hline
\end{tabular}
\end{adjustbox}

\label{tab:flops_segregation}
\end{table}
A substantial portion of the total FLOPs is consumed by classical layers and encoding. For instance, at a feature size of $110$, the classical and encoding components in BEL-based hybrid model consumes $3901$ FLOPs, representing the majority of the $4797$ total FLOPs, whereas, the quantum layer only requires $896$ FLOPs, about $18.7\%$ of the total.
On the other hand, for more sophisticated HQNN designs, particularly those employing SEL, the FLOPs consumption of the quantum layer remains constant, demonstrating the efficient handling of increasingly complex problems by more expressive quantum networks. This stability underscores the capability of these quantum layers to adapt to problem complexity without a proportional increase in computational demands.

\section{Conclusion} 
In this paper, we conducted a comprehensive benchmarking study to evaluate the computational efficiency of Hybrid Quantum Neural Networks (HQNNs) compared to classical neural networks (NNs). Our key findings aim at providing valuable insights to the research questions raised in the introduction as summarized below:

\textbf{(A1)}
Since there exists no definitive measure of computational complexity in NNs, relying solely on a single metric to draw conclusions about computational complexity may lead to limited insights.
Therefore, we employed two independent metrics, i.e., floating-point operations (FLOPs) and parameter count to evaluate computational complexity. We argue that if the trends observed in the results of these distinct metrics align, the conclusions derived from the analysis are more robust and meaningful.

\textbf{(A2)}
Our analysis reveals that as the problem complexity increases, classical NNs demonstrate a substantial rise in both FLOPs and the number of parameters due to the requirement of more sophisticated architecture to maintain a certain accuracy. This indicates limited adaptability of classical NNs to the increasing complexity of the problem. 
In contrast, HQNNs exhibit a more efficient computational footprint. 
Specifically, quantum layers within HQNNs consume significantly fewer FLOPs and require fewer parameters as problem complexity grows, compared to classical models. Additionally, HQNNs adapt well to the growing problem complexity since the rate of increase in FLOPs and parameter count (while going from low to high complexity of the problem) is significantly lower in HQNNs compared to classical models. These findings indicate that HQNNs exhibit superior adaptability to complex problem scenarios, highlighting their potential advantages over classical models in managing computational complexity.




%

\textbf{(A3)}
Given that both independent measures of complexity, i.e., FLOPs and parameter count demonstrate better adaptability in HQNNs to problem complexity over classical networks, we posit that there is strong evidence suggesting that the inclusion of quantum layers offers a computational advantage in NNs. We also advocate that further investigations using additional complexity measures are needed to fully understand the QML benefits over classical ML. Thus, we leave this question partially open for future research to explore with a broader set of metrics.

In summary, HQNNs offer a scalable, resource-efficient alternative to classical models, positioning them as a promising solution for complex tasks in machine learning.


\end{spacing}

\section*{Acknowledgements}
\vspace{-0.14cm}
This work was supported in part by the NYUAD Center for Quantum and
Topological Systems (CQTS), funded by Tamkeen under the NYUAD Research
Institute grant CG008.

\bibliographystyle{IEEEtran}
\bibliography{main.bib}

\end{document}